\documentclass[aps,prx,twocolumn,reprint,superscriptaddress]{revtex4-2}
\usepackage{physics}
\usepackage{braket}
\usepackage{graphicx}% Include figure files
\usepackage[hyperfootnotes=false]{hyperref}
\usepackage{color}
\usepackage{amssymb}
\definecolor{darkblue}{rgb}{0,0,0.5}
\hypersetup{
    colorlinks=true,
    linkcolor=black,
    filecolor=blue,
    citecolor=darkblue,
    urlcolor=black,
}

\newcommand{\jj}[1]{{{\textcolor{black}{#1}}}}

\begin{document}

%\preprint{}

%\title{Chaos Emerge with Exceptional Points in Reset-Driven Floquet Dynamics}
\title{\jj{Emergence of chaos with exceptional points in reset-driven Floquet dynamics}}

\author{Jia-Jin Feng}
\email{jiajinfe@usc.edu}
\affiliation{
Ming Hsieh Department of Electrical and Computer Engineering, University of Southern California, Los
Angeles, California 90089, USA
}
%{Your e-mail address}
%\thanks{}

\author{Quntao Zhuang}\email{qzhuang@usc.edu}
\affiliation{
Ming Hsieh Department of Electrical and Computer Engineering, University of Southern California, Los
Angeles, California 90089, USA
}
\affiliation{Department of Physics and Astronomy, University of Southern California, Los Angeles, California 90089, USA}


\date{\today}

\begin{abstract}
We investigate the spectral structure of reset-driven Floquet quantum channels generated by the Hamiltonian evolution of a many-body system followed by periodic resetting of a bath. 
\jj{
By tuning a perturbation parameter in the interacting Aubry–André–Harper(AAH) model, we find that exceptional points could emerge in the spectral transition from a symmetry-constrained ergodic regime to a fully chaotic regime. Across this transition, increasing the perturbation causes the real eigenvalues of the channel to drift, coalesce at exceptional points, and bifurcate into complex-conjugate pairs, reflecting the progressive breaking of symmetry constraints in operator space.
}
We further show that the channel spectrum sharply distinguishes chaotic, ergodic, many-body localized, and scarred dynamical regimes. Finally, we connect the leading channel eigenvalues to experimentally accessible probes based on quantum mutual information, establishing a link between the spectral organization of reset-driven quantum channels and observable relaxation dynamics. 
\jj{
We implement the protocol on the IBM Quantum Platform and verify that the dynamical differences are reflected in the decay of mutual information.
}
\end{abstract}

%\keywords{}
% We find that the spectral structure of the chan-
% nel differs qualitatively across chaotic, ergodic, and MBL
% regimes
\maketitle

\section{Introduction}

Exceptional points (EPs) are singular degeneracies of non-Hermitian operators at which both eigenvalues and eigenvectors coalesce \cite{choi2018,Luitz2019,el2018non,Huang2025,luo2018,Xue2026,Xu2026,Mohammad2019}. At an EP, the generator becomes defective, giving rise to nonanalytic spectral behavior and qualitatively modified dynamics. EPs have been studied extensively in non-Hermitian Hamiltonians, with a paradigmatic example provided by the $\mathcal{PT}$-symmetric two-level Hamiltonian $H_{\rm PT}=iJ_{z}\sigma_z+J_{x}\sigma_x$.
When the gain-and-loss term $J_z$ dominates, the eigenvalues $E$ become purely imaginary, so that the time-evolution factor $e^{-iEt/\hbar}$ is purely real and the dynamics is governed by exponential amplification or decay rather than oscillation. By contrast, when the coherent coupling $J_x$ dominates, the corresponding time-evolution factor $e^{iEt/\hbar}$ is complex and describes oscillation.
The transition between these two regimes occurs at an EP.

More recently, non-Hermitian physics has been extended to open quantum systems described by master equations, where the Liouvillian spectrum governs continuous-time dissipative dynamics~\cite{Minganti2019,sun2024,wu2026,Hanai2020,Zlatko2025,xue2025,Chaduteau2026}. By contrast, discrete-time open-system dynamics are naturally described by quantum channels~\cite{Ernesto2026,Wong2026,chen2026}, namely completely positive trace-preserving(CPTP) maps. This framework is particularly relevant for protocols involving strong measurements, resets, and feedback operations, which are intrinsically stroboscopic and therefore most naturally formulated at the level of quantum channels.
Despite its importance, the non-Hermitian spectral structure of many-body quantum channels remains much less explored, because their spectra are highly complex.
In particular, the eigenvalues of a quantum channel directly determine relaxation rates, memory effects, and information retention.
EPs can therefore play a central role by reorganizing the channel spectrum and qualitatively modifying the resulting dynamical behavior.

Modern dynamic-circuit platforms, central to quantum error correction, quantum algorithms~\cite{hu2024overcoming,zhang2024generative} and information processing~\cite{Baumer2024,Langbehn2024,langbehn2025,zhang2025holographic}, naturally realize such channels.
In these architectures, coherent evolution is interspersed with mid-circuit measurements and conditional resets, so that the reduced system evolves under a quantum channel.
Unlike effective non-Hermitian Hamiltonian descriptions, which may violate probability conservation unless supplemented by additional prescriptions~\cite{Moiseyev_2011}, reset-induced channels provide a physically consistent framework in which non-unitarity arises from tracing out and resetting degrees of freedom \cite{yokoyama2026}.
%The resulting superoperator spectrum therefore encodes genuine open-system dynamics without ad hoc modifications.

\begin{figure*}[tb]
\includegraphics[clip = true, width =2\columnwidth]{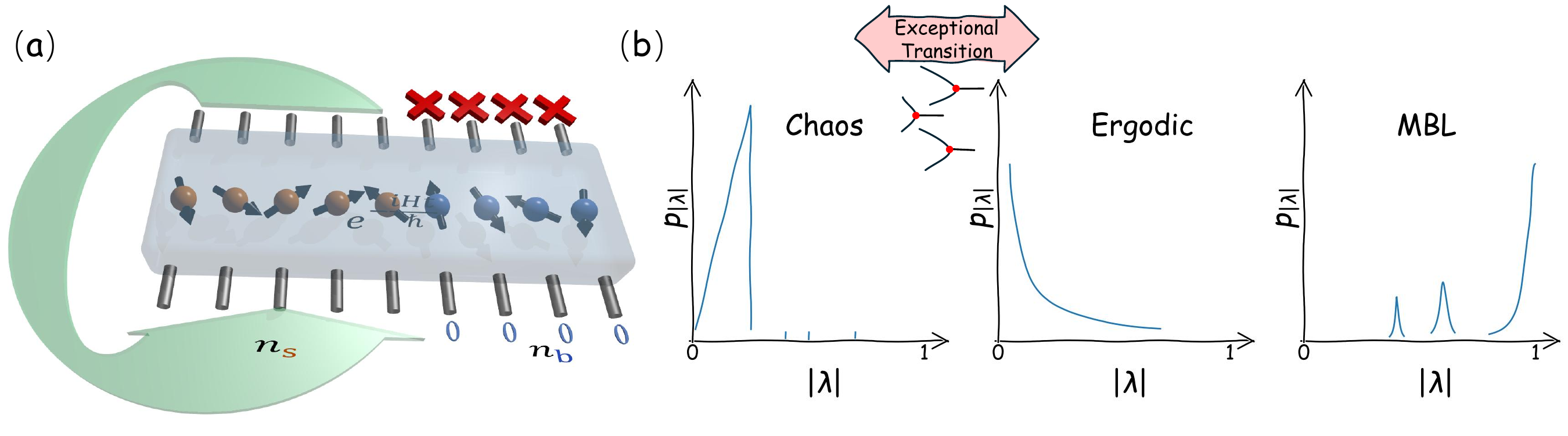}
\caption{\label{fig:K}(a) Circuit diagram illustrating the superoperator $\mathcal{K}$. The combined system and bath evolve under the Hamiltonian, after which the bath is measured and reinitialized. The resulting system output is then used as the input for the subsequent iteration. (b) Typical probability distributions(illustrative sketch, not numerical data) of the eigenvalues of $\mathcal{K}$ for different underlying Hamiltonian regimes.
Upon tuning the Hamiltonian from the chaotic to the ergodic regime, the channel spectrum undergoes an EP-mediated transition.}
\end{figure*}
%\section{Overview}

In this work we investigate \emph{reset-driven Floquet} channels, generated by periodically evolving a finite-size system and bath under an interacting many-body Hamiltonian and resetting the bath after each cycle, as illustrated in Fig.~\ref{fig:K}(a).
\jj{Focusing on the interacting Aubry–André–Harper(AAH) model, we discover that tuning a perturbative parameter drives an EP-related spectral transition between symmetry-constrained ergodic dynamics and fully chaotic dynamics~\cite{Zhang2016,jiang2018,Kong2024,Salvatore2021}.
In the ergodic regime, the channel spectrum contains many real eigenvalues with large magnitudes.
As perturbation increases, eigenvalues drift along the real axis, coalesce at EPs, and bifurcate into complex-conjugate pairs with the characteristic square-root splitting.
Upon further tuning, additional symmetry breaking occurs \cite{Friederike2025,Lin2026} and global chaos emerges, accompanied by a proliferation of EPs across the spectrum.
}

Beyond identifying individual EPs, we find that the channel spectrum itself provides a diagnostic of distinct dynamical regimes as sketched in Fig. \ref{fig:K}(b). In the strongly chaotic regime, where the Hamiltonian breaks most symmetries, the channel spectrum approaches random-matrix behavior~\cite{Salvatore2021,mok2026,Wang2022} and is well described by the circular law, with eigenvalues distributed approximately uniformly inside a disk in the complex plane. In the ergodic regime, by contrast, global symmetries constrain the dynamics, and the channel spectrum forms a smooth, continuous distribution. In the many-body localization(MBL) regime, the presence of an extensive set of quasi-local integrals of motion further constrains the dynamics. Consequently, the channel spectrum is discrete, with eigenvalues clustered near unity, reflecting long-lived modes and slow relaxation. We further identify a distinct cluster of modes whose moduli are close to 1, suggesting a possible spectral signature of discrete time-crystalline behavior.
For systems with weak eigenstate thermalization hypothesis(ETH) breaking, in which chaotic dynamics coexist with anomalously long-lived modes such as quantum many-body scars~\cite{Schecter2019}, the channel spectrum develops a hybrid structure. It retains the broad spectral features characteristic of chaotic dynamics, while also exhibiting isolated, weakly decaying modes associated with the scarred subspace.

%\QZ{We} emphasize that the results presented here are obtained for a specific family of reset-driven Floquet channels and Hamiltonian models. We do not claim a universal classification of many-body dynamics from channel spectra. Instead, we identify spectral signatures that distinguish the regimes realized in these models. In particular, the parameter Jxxx
%​simultaneously breaks U1 symmetry and introduces nonintegrability,therefore, the observed exceptional-point proliferation should be interpreted as accompanying the combined lifting of symmetry constraints and the onset of full-Hilbert-space chaos, rather than as a unique signature of chaos alone.

\jj{
Finally, we relate the leading eigenvalues of the channel to experimentally accessible probes based on the quantum mutual information, thereby establishing a direct connection between the spectral structure relative to the EP and the observable relaxation dynamics. We implement the protocol on the IBM Quantum Platform and investigate the resulting differences in the dynamical behavior of the mutual information.
}

\section{Setup}

The protocol of the reset-driven Floquet channel is shown in Fig.~\ref{fig:K}. We consider a  finite-size composite system comprising $n_{\rm s}$ system qubits and $n_{\rm b}$ bath qubits, with a total of $n_{H}=n_{\rm s}+n_{\rm b}$ qubits.
The Hilbert-space dimension of the system is $N_{\rm s}=2^{n_{\rm s}}$ and the bath is $N_{\rm b}=2^{n_{\rm b}}$.
For simplicity, the bath is prepared in the product state $|0\rangle^{\otimes n_{\rm b}}$, while the system is initialized in an arbitrary state. The composite system undergoes Hamiltonian evolution generated by $H$ for a duration $t$. Following the evolution, the bath qubits are measured and reset, while the output state of the system is recycled as the input to the next round. Consequently, the evolution of the system density matrix in each iteration is governed by the superoperator
\begin{eqnarray}
    \mathcal{K}\left(\rho_{\rm s}\right)&=&{\rm Tr}_{\rm b}\left[ e^{-i\frac{Ht}{\hbar}}  \left(\rho_{\rm s}\otimes \rho_{\rm b0} \right) e^{i\frac{Ht}{\hbar}} \right] \nonumber \\
    &=&\sum_{m=0}^{N_{\rm b}-1} \langle m|  e^{-i\frac{Ht}{\hbar}}|0\rangle_{\rm b}  \rho_{\rm s}\langle 0| e^{i\frac{Ht}{\hbar}}|m\rangle_{\rm b} ~,
\label{eq:Kraus}
\end{eqnarray}
where we select $\rho_{\rm b0}=|0\rangle\langle0|^{\otimes n_{\rm b}}$ for convenience.
$\{|m\rangle_{\rm b}\}$ denotes the measurement basis of the bath Hilbert space.
The evolution time $t$ is taken to be sufficiently long so that the dynamics become fully scrambled \cite{Liang2025,Wang2022}; otherwise, frequent reset operations can drive the system into a quantum-Zeno regime \cite{Feng2026,kumar2026}.
Since $\mathcal{K}$ is a linear map, the density matrix can be vectorized, allowing $\mathcal{K}$ to be represented as a large matrix acting on an $N_{\rm s}^2$-dimensional space. Its spectral properties can then be analyzed directly \cite{zhang2025scaling,Zlatko2025}.

% \QZ{For example, consider the Pauli basis, the matrix $K_{\mu \nu}=\frac{1}{d?}\tr\left(\sigma_\mu \mathcal{K}(\sigma_\nu)\right)$ is real because $\sigma_nu$ and $\mathcal{K}(\sigma_\nu)$ are both real. However, $K_{\nu\mu}^\star=\tr\left(\sigma_\nu \mathcal{K}(\sigma_\mu)\right)$ generally is not equal to $K_{\mu \nu}$. So the matrix is not Hermitian. Real matrices like $K_{\mu \nu}$ have eigenvalues in conjugate pairs $a\pm b i$, with corresponding eigenvectors $\sigma$ and $\sigma^\star$ as well. Such pairing structure sometimes becomes degenerate, when the imaginary part goes to zero, which is referred to as an exceptional point. With exception points absent, the matrix becomes defective and Eq.~\eqref{Kn_decomposition} requires corrections.}

Away from EPs, where the channel $\mathcal{K}$ is nonsingular and diagonalizable, the initial density matrix can be decomposed in the eigenoperator basis of $\mathcal{K}$ as
\begin{eqnarray}
    \rho_{\rm s0}=\sum_{m=0}^{N_{\rm s}^2-1}c_m \varrho_m ~,
\end{eqnarray}
where $\varrho_m$ are the right eigenoperators of $\mathcal{K}$. The left eigenoperator with the conjugate eigenvalue is $\mu_m$. Then the coefficient is $c_m={\rm Tr}(\mu_m^\dagger\rho_{\rm s0})$, with the normalization ${\rm Tr}(\varrho_m^\dagger\varrho_{m})=1$ and biorthonormalization condition ${\rm Tr}(\mu_m^\dagger\varrho_{m'})=\delta_{m,m'}$ \cite{Minganti2019}. Repeating the process $n_{\rm r}$ times yields the non-unitary evolution
\begin{eqnarray}
    \mathcal{K}^{n_{\rm r}}\left(\rho_{\rm s0}\right)=\sum_{m=0}^{N_{\rm s}^2-1}\lambda_m^{n_{\rm r}} c_m \varrho_m ~,
    \label{eq:Kn}
\end{eqnarray}
where $\lambda_m$ denotes the eigenvalue associated with $\varrho_m$. Consequently, the dynamical properties of the evolution are governed by the spectral structure of $\mathcal{K}$. If an eigenvalue $\lambda_m$ is complex, there exists another index $l$ such that $\lambda_l=\lambda_m^{*}$. Complex eigenvalues therefore occur in conjugate pairs, ensuring that the density matrix remains Hermitian. At the EPs, the situation is more complicated which can be seen in the Appendix \ref{app:EP}.

We take $H$ to be an interacting AAH spin chain \cite{Marko2018,Yoo2020,Chen2025}
\begin{eqnarray}
H_{\rm AAH}&=&\sum_{m=0}^{n_{H}-2} \left( J_{xx} \sigma_m^x \sigma_{m+1}^x+J_{yy} \sigma_m^y\sigma_{m+1}^y+J_{zz}\sigma_m^z\sigma_{m+1}^z\right)\nonumber\\
&&+J_z\sum_{m=0}^{n_{H}-1}\cos(\omega m) \sigma_m^z ~,
\label{eq:AAH}
\end{eqnarray}
where $\omega$ is irrational, so the last term realizes a pseudo-randomness on-site potential. $J_{xx}=J_{yy}$ is set to the same. For small $J_z$ the model is ergodic, while for large $J_z$ it exhibits MBL. The phase transition point is near $J_z/J_{xx}=2$, which is rounded by interactions \cite{Marko2018} and finite-size effects \cite{sierant2025many,BREZIN1985,jiang2026}. To eliminate remaining conservation such as $U(1)$ symmetry, we introduce a three body interaction term to promote chaos across the full Hilbert space as,
\begin{eqnarray}
    H_{XXX}&=H_{\rm AAH}+J_{xxx}\sum_{m=1}^{n_{H}-2}\sigma_{m-1}^x\sigma_m^x \sigma_{m+1}^x ~.
\end{eqnarray}
In the following, we study how the Kraus dynamics generated by Eq.~\ref{eq:Kraus} depends on the underlying Hamiltonian, contrasting qualitatively distinct regimes.

\section{Global Chaos}
\begin{figure}[tb]
\includegraphics[clip = true, width =\columnwidth]{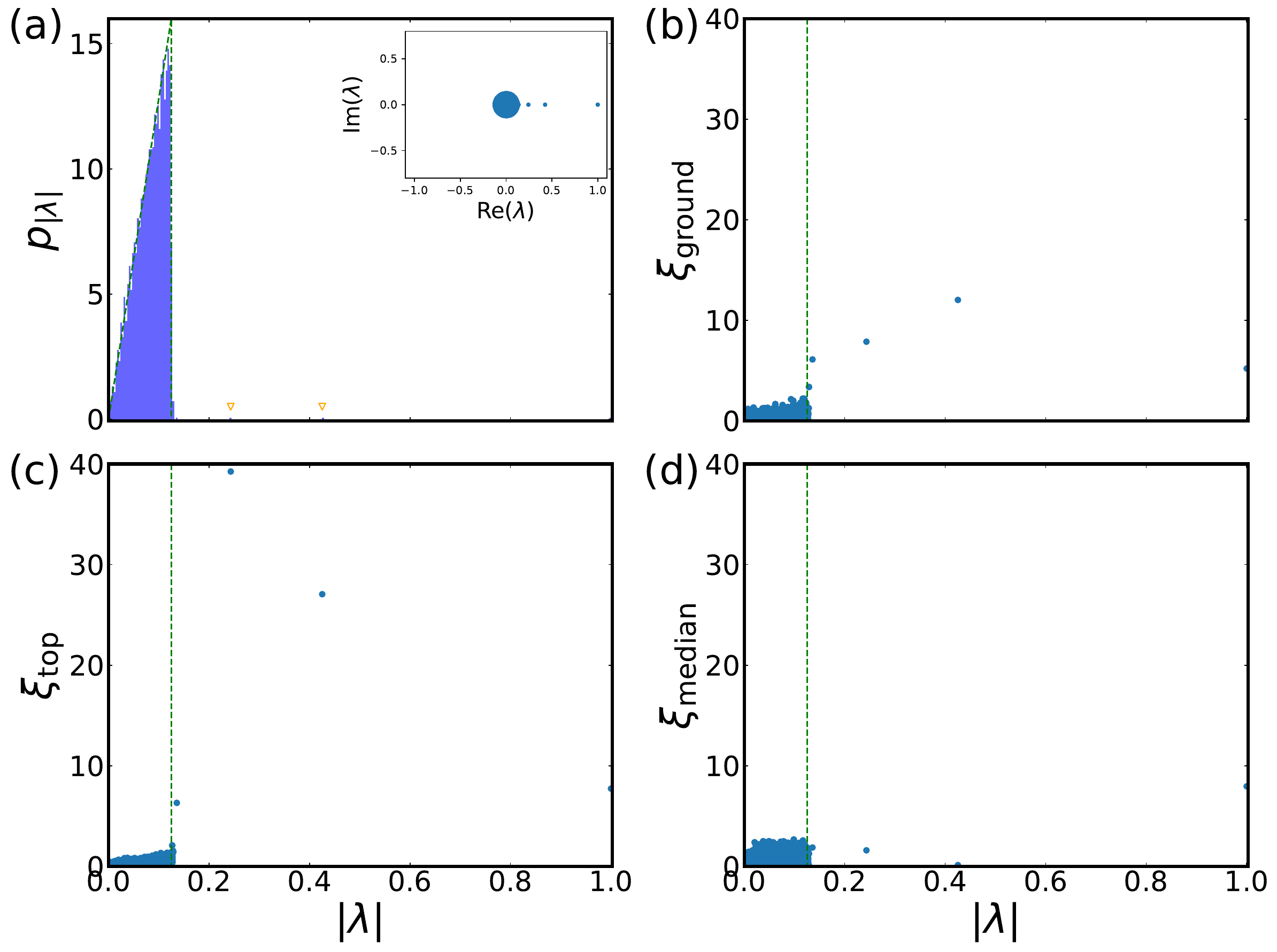}% Here is how to import EPS art
\caption{\label{fig:EigXXX} (a) Probability distribution of the eigenvalues of the superoperator $\mathcal{K}$. The green dashed line denotes the triangular distribution expected for Haar-random dynamics. (b–d) Rescaled overlaps between eigenstates of $\mathcal{K}$ and the (b) ground state, (c) top state, and (d) median state of the Hamiltonian $H$. Each point corresponds to a single eigenstate of $\mathcal{K}$. The green dashed line marks the reference value $1/\sqrt{N_{\rm b}}$. Other parameters are $J_{zz}=J_z=0.1J_{xx}$, $J_{xxx}=2$, $t=100\hbar/J_{xx}$, $n_{\rm s}=n_{\rm b}=6$
}
\end{figure}

\begin{figure}[tb]
\includegraphics[clip = true, width =\columnwidth]{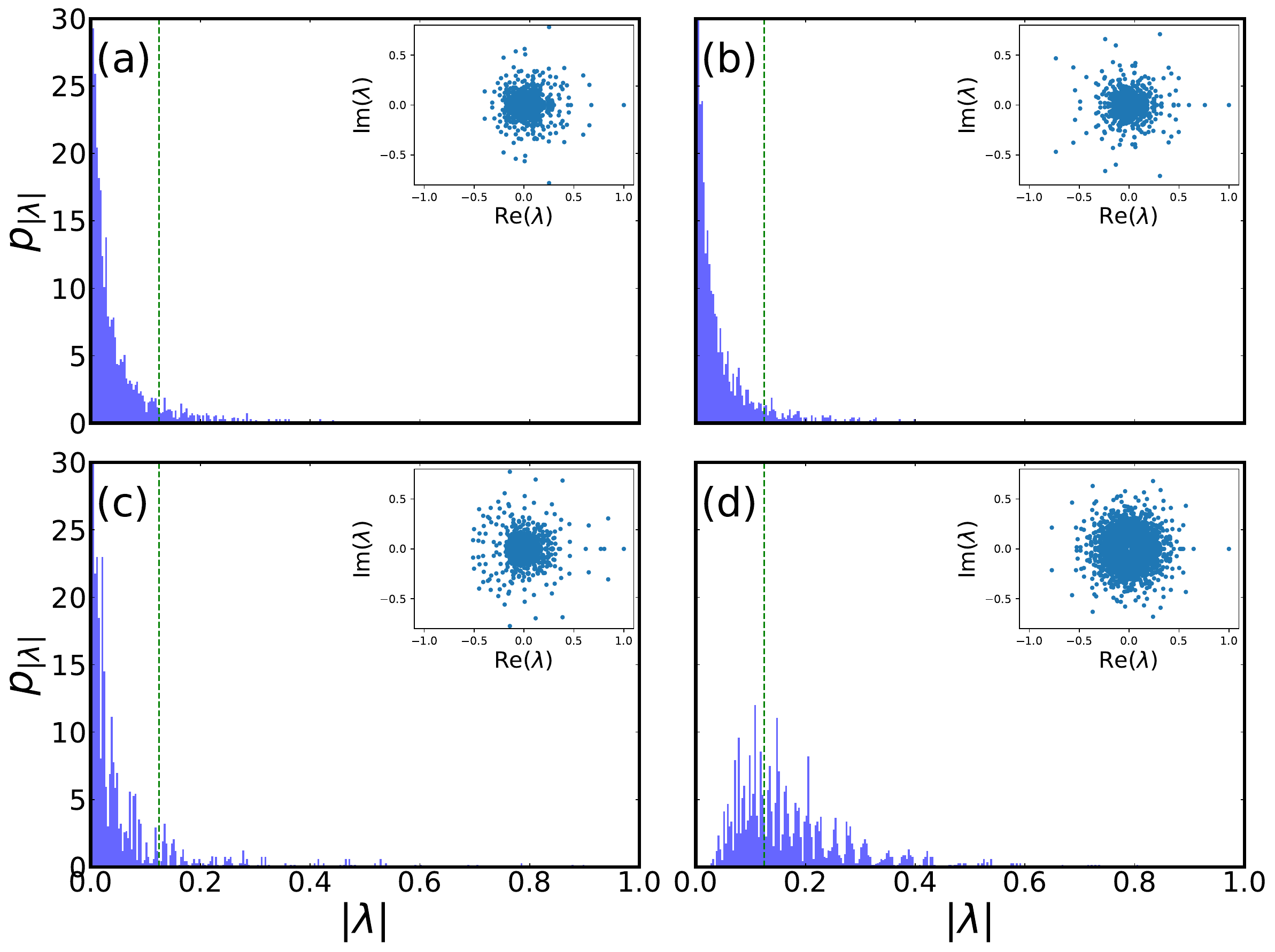}
\caption{\label{fig:Ergodic} Probability distribution of the eigenvalues of $\mathcal{K}$.
The green dashed line indicates the threshold $1/\sqrt{N_{\rm b}}$.
Parameters are $n_{\rm s}=6$, $n_{\rm b}=10$, $J_{xxx}=0$, and $J_z=0.1J_{xx}$.
Panels (a)(b) correspond to the interacting case $J_{zz}=0.3J_{xx}$, while panels (c)(d) show the noninteracting case $J_{zz}=0$.
The evolution times are $t=100\hbar/J_{xx}$ for (a)(c) and $t=200\hbar/J_{xx}$ for (b)(d).
}
\end{figure}

\jj{Chaotic Hamiltonians can be described using random matrix theory (see Appendix \ref{app:CE}).
For the strongly chaotic Hamiltonian $H_{XXX}$, the unitary evolution $e^{-iHt/\hbar}$ possesses few relevant global symmetries beyond energy conservation.} The probability distribution of the eigenvalues of the superoperator $\mathcal{K}$ is shown by the blue histogram in Fig.~\ref{fig:EigXXX}(a). The unitary evolution $e^{-iHt/\hbar}$ in Eq.~(\ref{eq:Kraus}) can be approximated by a Haar-random unitary, for which the eigenvalue spectrum follows the circular law as shown in the inset of Fig.~\ref{fig:Ergodic}(a). When projected onto the absolute-value axis, this corresponds to a triangular distribution, shown as the green dashed line in Fig.~\ref{fig:EigXXX}(a). Such random evolution generically leads to an exponential decay of information and local observables with the number of rounds $n_\mathcal{K}$~\cite{zhang2025scaling,Morningstar2022}.

Despite this overall agreement, we observe a small number of outliers with large eigenvalues that deviate from the circular law. These outliers satisfy $|\lambda|>1/\sqrt{N_{\rm b}}$ and lie on the real axis. To characterize their structure, we define a rescaled overlap between the corresponding eigenmodes of $\mathcal{K}$ and eigenstates of the Hamiltonian,
\begin{eqnarray}
    \xi=N_{\rm s} | {\rm Tr} (\rho_\Psi \varrho_m) |
\end{eqnarray}
where $\rho_\Psi =\langle0_{\rm b}|\Psi\rangle\langle\Psi|0_{\rm b}\rangle/{\rm Tr}(\langle0_{\rm b}|\Psi\rangle\langle\Psi|0_{\rm b}\rangle)$ and $|\Psi\rangle$ is an eigenstate of the full Hamiltonian $H$. On average, most eigenmodes are distributed around $\xi\approx 1$, consistent with Haar-random behavior. In contrast, the outliers exhibit significantly enhanced overlap with the ground state and the top(highest-energy) state of $H$, as shown in Fig.~\ref{fig:EigXXX}(b)(c), while their overlap with the median(-energy) state remains small [Fig.~\ref{fig:EigXXX}(d)]. For spatially local Hamiltonians, the median state typically obeys a volume-law entanglement entropy and thus resemble Haar-random states. By contrast, eigenstates near the Hamiltonian spectral edges exhibit reduced entanglement and violate the volume law~\cite{kuwahara2020area,Brydges2019}. We therefore attribute the breakdown of the circular law for these outliers to their strong overlap with low- and high-energy eigenstates of the Hamiltonian.

\section{Symmetry-Constrained Ergodicity}
\jj{Here we consider the case of an ergodic Hamiltonian that cannot be described by random-matrix theory over the full Hilbert space (see Appendix \ref{app:CE}).}
In the case of $J_{xxx}=0$, a $U(1)$ symmetry for the Hamiltonian emerges, such that the total spin $S_z=\sum_m\sigma_m^z$ is conserved by the Hamiltonian. Consequently, the Hamiltonian dynamics is no longer chaotic in the full Hilbert space, and the Haar-random approximation over the entire Hilbert space is not applicable. Although the Hamiltonian commutes with the total spin, $[H,S_z]=0$, the superoperator does not preserve this symmetry, since $S_z\mathcal{K}(\rho)S_z-\mathcal{K}(S_z \rho S_z)\neq0$.
Thus, the channel dynamics cannot be confined to a single fixed-$S_z$ sector, and the full system Hilbert space must be retained.
Nevertheless, the conservation law of the joint system imposes a residual structure on the reduced dynamics. Because the bath is reset to the $S_z=-n_{\rm b}$ sector after each round, magnetization can only be transferred from the system to the bath, leading to a monotonic decrease of the system magnetization. Consequently, after vectorizing the density matrix, the matrix representation of $\mathcal{K}$ acquires a block-triangular form in the magnetization-resolved basis~\cite{Li2026}.
We further find that, when $H_{\rm AAH}$ is ergodic, the resulting channel possesses many large-magnitude real eigenvalues, as shown in the inset of Fig.~\ref{fig:Ergodic}.

In the low-symmetry regime with a finite interaction term $J_{zz}\neq 0$, the spectrum is only weakly dependent on the evolution time once it is sufficiently long, as revivals are strongly suppressed. Consequently, Fig.~\ref{fig:Ergodic}(a) closely resembles Fig.~\ref{fig:Ergodic}(b). The probability distribution is predominantly concentrated near $|\lambda|=0$, with an exponentially decaying tail. This behavior originates from the sparsity of the superoperator, which can be approximately described by a finite-connectivity graph matrix \cite{Kuhn_2008,Metz_2019}. Notably, the tail extends beyond the threshold $1/\sqrt{N_{\rm b}}$ (indicated by the green dashed line in Fig.~\ref{fig:Ergodic}), with a substantial weight. This feature qualitatively distinguishes the dynamics from that of a fully chaotic Hamiltonian acting on the entire Hilbert space.

\begin{figure}[tb]
\includegraphics[clip = true, width =\columnwidth]{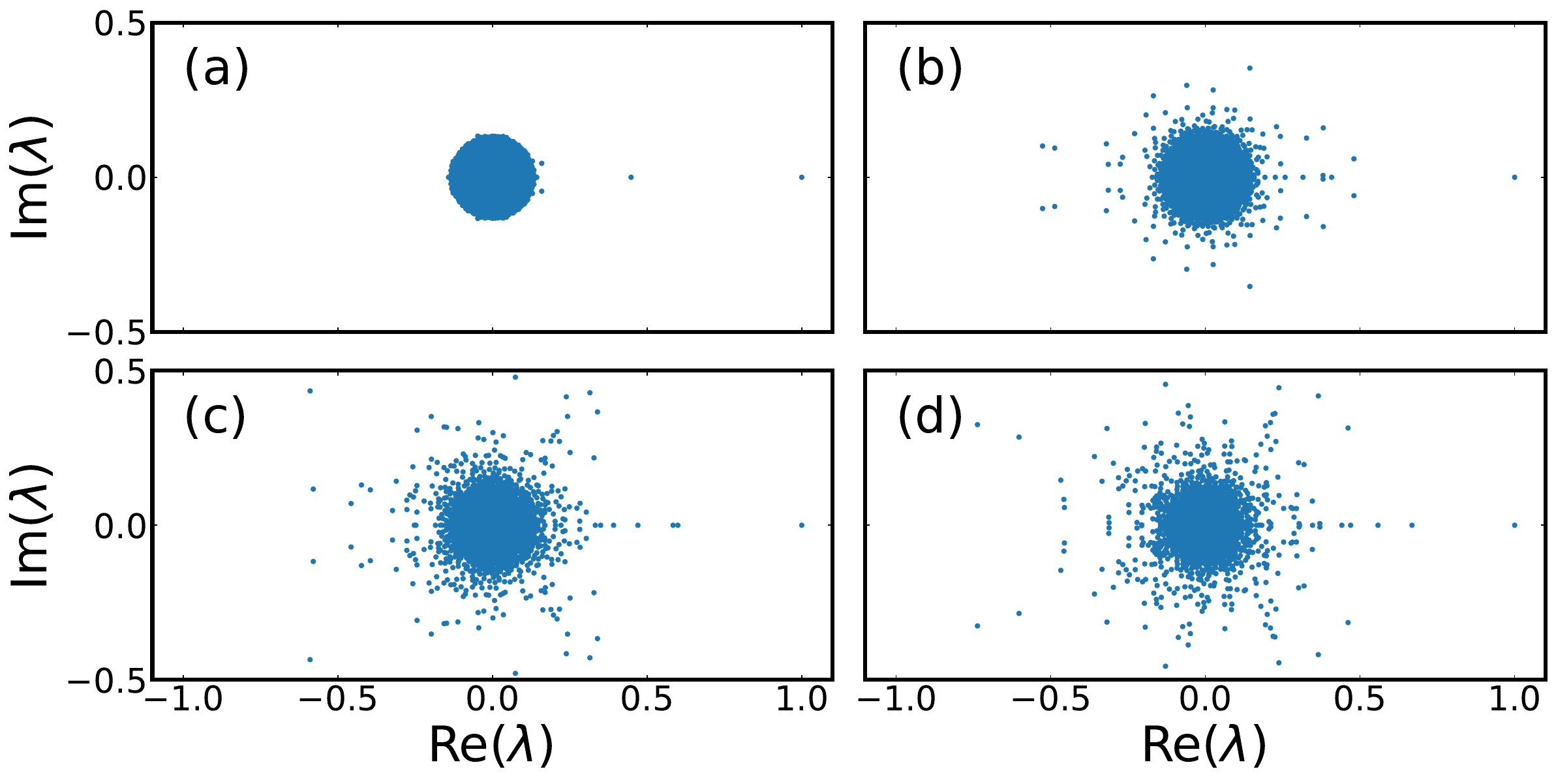}
\caption{\label{fig:TC2} Eigenvalues of $\mathcal{K}$. Each dot denotes one eigenvalue in the complex plane for (a) $J_{xxx}=0.1J_{xx}$, (b) $J_{xxx}=0.05J_{xx}$, (c) $J_{xxx}=0.02J_{xx}$, and (d) $J_{xxx}=0.01J_{xx}$. Other parameters are $J_{zz}=J_z=0.1J_{xx}$, $t=100\hbar/J_{xx}$, and $n_{\rm s}=n_{\rm b}=6$.
}
\end{figure}

In contrast, in the high-symmetry regime without interactions $J_{zz}= 0$, the spectrum is highly sensitive to the evolution time due to pronounced revivals. As a result, Fig.~\ref{fig:Ergodic}(c) differs markedly from Fig.~\ref{fig:Ergodic}(d). At certain evolution times, the distribution is strongly peaked near $|\lambda|=0$, as shown in Fig.~\ref{fig:Ergodic}(c), while at other times it shifts away from zero, as illustrated in Fig.~\ref{fig:Ergodic}(d). In both cases, the spectrum retains a tail extending beyond $1/\sqrt{N_{\rm b}}$.

% The spectrum is discrete \cite{Zhang2016}.
% The energy levels are protected by global symmetry. The oscillation will be forever. It is sensitive to time

% finite size effect make local symmetry to global symmetry

% quasisymmetry\cite{Ren2021}

% locate near 0 \cite{Kuhn_2008,Metz_2019}

\section{Exceptional Points Proliferation at the Emergence of Chaos}

In this section, we investigate the quantum transition from ergodic to chaotic dynamics. Fig.~\ref{fig:TC2} shows how the channel eigenvalues evolve as the perturbation parameter is tuned. As this parameter is reduced, the spectrum deforms from a circular-law distribution toward a broadened distribution, with an increasing number of eigenvalues accumulating along the real axis.

\begin{figure}[tb]
\includegraphics[clip = true, width =\columnwidth]{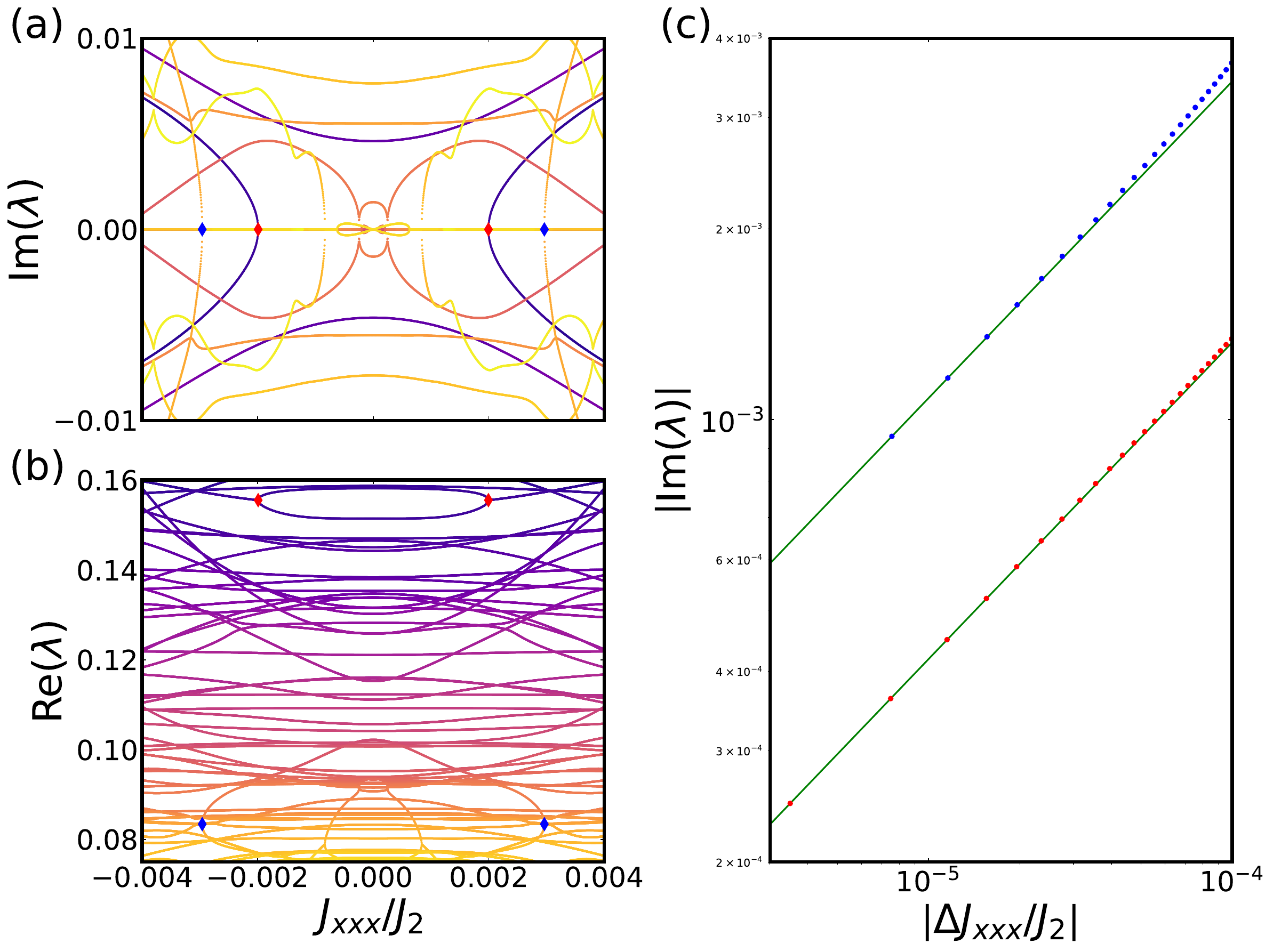}
\caption{\label{fig:EP}
Selected eigenvalue bands of $\mathcal{K}$ with large positive real parts. (a) Imaginary and (b) real parts of the eigenvalues as a function of the perturbation parameter. Prisms mark representative EPs. (c) Log–log plot of the eigenvalue magnitude versus the perturbation parameter; dots highlight values in the vicinity of the EPs indicated in (a). The green line shows the $\sim\sqrt{\Delta J_{xxx}}$ scaling. 
}
\end{figure}

For an ergodic channel, many of the eigenvalues $\lambda$ are real and large. Figs.~\ref{fig:EP}(a) shows that the imaginary parts of some eigenvalues vanish at $J_{xxx}=0$. As the perturbation parameter is increased, the eigenvalues drift along the real axis, as illustrated in Fig.~\ref{fig:EP}(b). This motion can bring two real eigenvalues together, leading to their coalescence and the formation of an EP. Representative EPs are highlighted by the prisms in Figs.~\ref{fig:EP}(a) and \ref{fig:EP}(b). Upon further increasing the perturbation parameter, the coalesced eigenvalue pair leaves the real axis and splits into a complex-conjugate pair, $\lambda_l=\lambda_m^*$, with opposite imaginary parts, as shown in Fig.~\ref{fig:EP}(a).

Near a second-order EP, the local spectral splitting obeys the characteristic square-root scaling, $|\Delta\lambda|\propto\sqrt{|\Delta J_{xxx}|}$. This behavior is verified in the log-log plot in Fig.~\ref{fig:EP}(c). As the perturbation parameter is increased further, EPs proliferate, and an increasing number of real eigenvalue pairs coalesce and split into complex-conjugate pairs. Crossing an EP signals the breaking of an underlying symmetry~\cite{Xiao2021}. Thus, as more EPs are encountered, more symmetry constraints are lifted, and the channel becomes progressively more chaotic.

\section{Many-Body Localization}

\begin{figure}[tb]
\includegraphics[clip = true, width =\columnwidth]{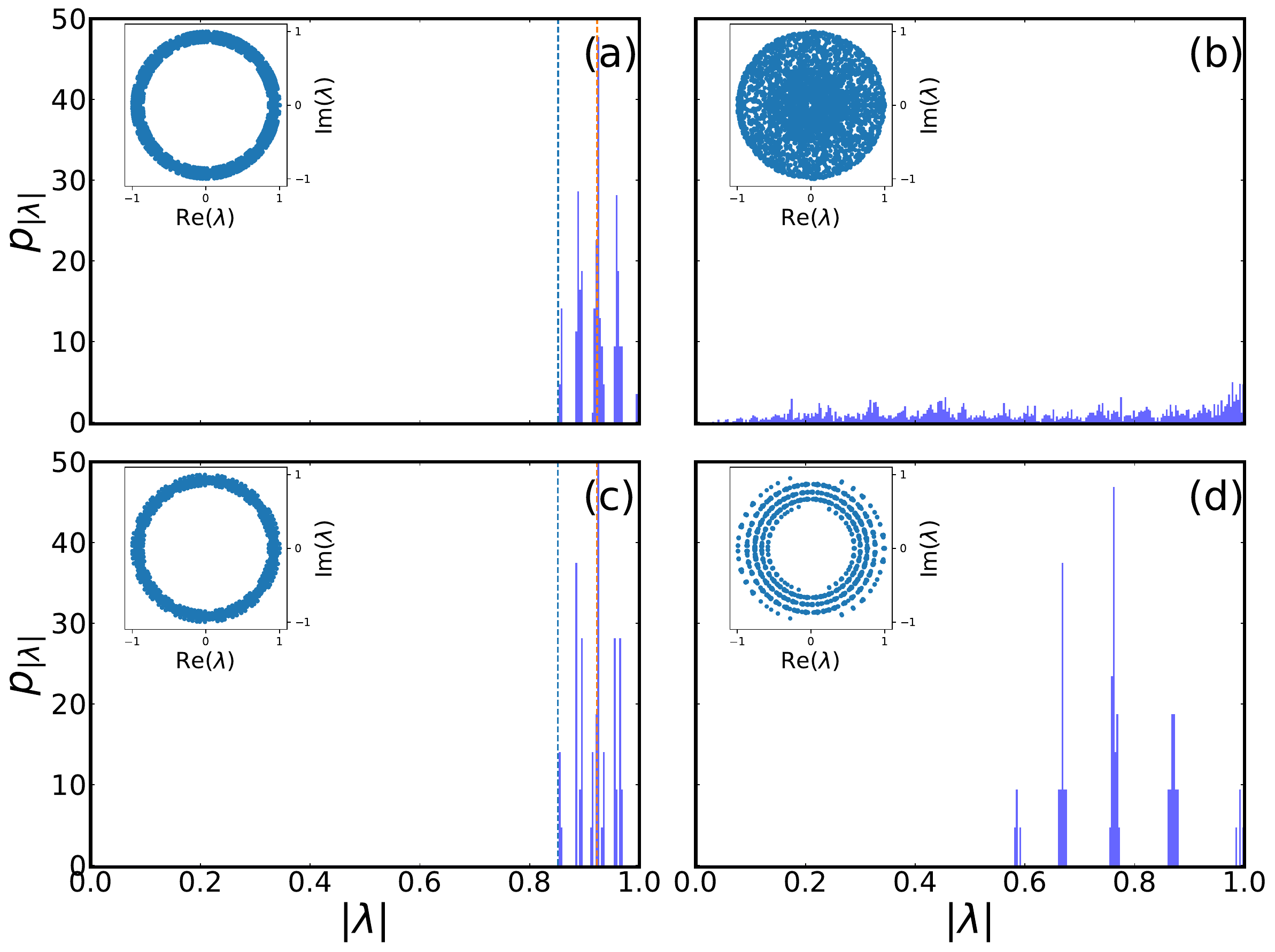}
\caption{\label{fig:MBL} Probability distributions of the eigenvalues of the superoperator $\mathcal{K}$.
The system contains $n_{\rm s}=6$ system qubits and $n_{\rm b}=10$ bath qubits, with $J_{xxx}=0$ and $J_z=5J_{xx}$. (a)(b) correspond to $J_{zz}=0$, while (c)(d) correspond to $J_{zz}=0.1$. The evolution time is $t=0.2\hbar/J_{xx}$ for (a)(c) and $t=200\hbar/J_{xx}$ for panels (b)(d).
}
\end{figure}

When the quasiperiodic potential $J_z$ is large, the system enters MBL \cite{LIU20253991}. In this regime, the dynamics is governed by an extensive set of quasi-local conserved quantities, known as local integrals of motion (LIOMs) \cite{jiang2025}, which form local orbits in Hilbert space. For sufficiently large systems, these LIOMs are nearly independent of the total system size. Since only LIOMs located near the system–bath boundary can effectively couple to the bath, the resulting leakage is strongly suppressed, leading to large eigenvalues $|\lambda|\gg1/\sqrt{N_{\rm b}}$, as shown in Fig.~\ref{fig:MBL}. Moreover, strong localization implies a small dispersion of LIOM energies, resulting in a highly discrete spectrum, also visible in Fig.~\ref{fig:MBL}.

Although MBL can strongly suppress the external coupling between the system and the bath, it simultaneously suppresses coherent transitions within the system itself. As a result, the ratio between the dissipation and the intrinsic transition does not change qualitatively. 
%Consequently, MBL alone does not drive the dynamics into a regime with complex spectral structure; instead, the evolution spectrum remains real.

The presence of strong local oscillations \cite{zhao2019} renders the dynamics sensitive to the evolution time. For short-time dynamics, only qubits near the boundary contribute appreciably, which significantly reduces the effective Hilbert-space dimension. The corresponding analytical result, shown by the dashed line in Fig.~\ref{fig:MBL}(c), is in good agreement with the numerical data. At longer times, small biases accumulate, preventing an exact analytical prediction even though the LIOM dispersion remains weak. Nevertheless, the spectrum remains discrete in Fig.~\ref{fig:MBL}(d), as the local oscillatory dynamics is preserved.

Introducing a finite interaction $J_{zz}$ destabilizes the MBL phase \cite{Marko2018}. In the short-time regime, the dynamics can still be well approximated by considering only the boundary qubit, as shown in Fig.~\ref{fig:MBL}(a). In contrast, at long times the sharp spectral peaks broaden and fragment into multiple smaller peaks, as illustrated in Fig.~\ref{fig:MBL}(b). Notably, the peak near $|\lambda|\approx 1$ is more robust against the interaction $J_{zz}$ than the others, retaining both its sharpness and position, while the remaining peaks exhibit substantial dispersion.

A cluster of eigenvalues with $|\lambda|\approx 1$ and $\lambda \neq 1$ appears in the channel spectrum, revealing a period-$n$ structure in the relaxation dynamics. This spectral feature is indicative of discrete time-crystalline behavior, in which the corresponding modes acquire a phase $2\pi/n$ upon one application of the Floquet channel and thus recur after $n$ periods~\cite{feng2025}. We further find that these eigenvalues remain robust against perturbations in $J_{zz}$, indicating that the period-$n$ response is not a fine-tuned effect. This observation is consistent with previous studies identifying many-body localization as a natural platform for stabilizing discrete time crystals~\cite{Liu2023,Randall2021}.

\begin{figure}[tb]
\includegraphics[clip = true, width =\columnwidth]{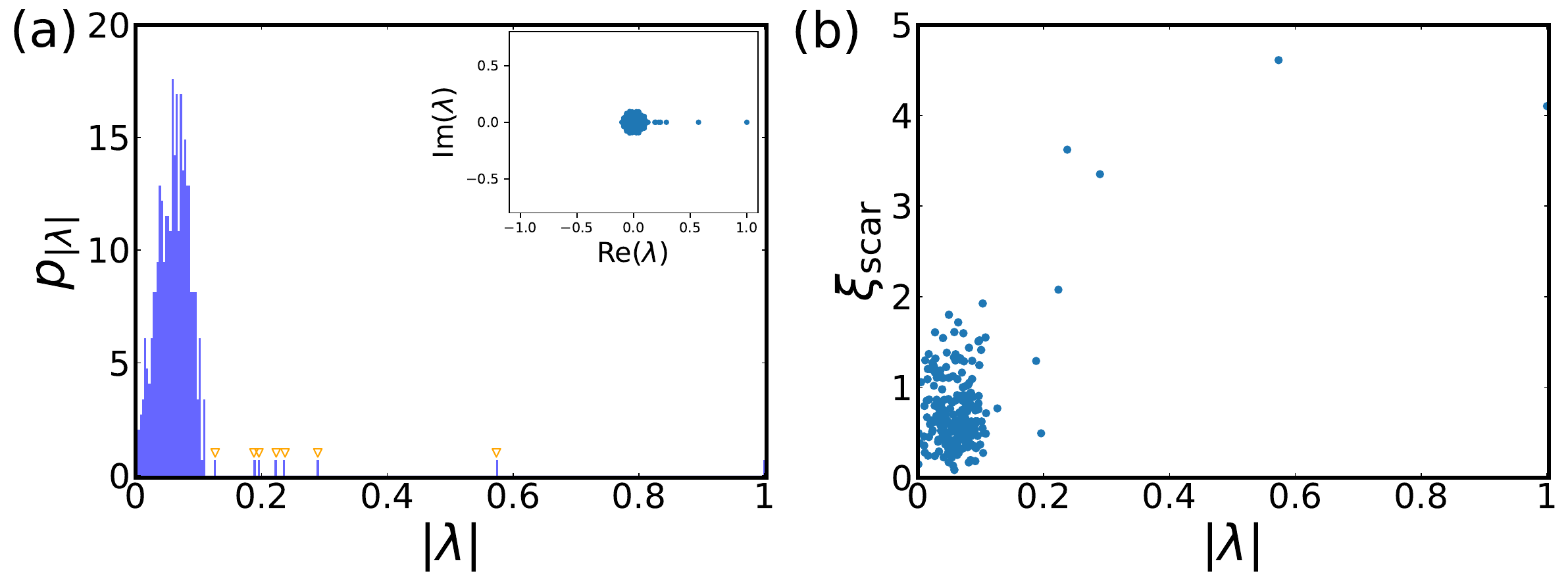}
\caption{\label{fig:PXP} (a) Probability distribution of the eigenvalues of the superoperator $\mathcal{K}$.
(b) Average rescaled overlap between the corresponding eigenmodes and the exact many-body scar states.
The system consists of $n_{\rm s}=8$ system qubits coupled to $n_{\rm b}=12$ bath qubits, with evolution time $t=200\hbar/J_{xx}\Omega$.
}
\end{figure}

\section{System Hosted Scar States}
In chaotic many-body systems, the majority of eigenstates are expected to satisfy ETH \cite{Kim2014, pilatowsky2025}. Nevertheless, ETH can be weakly violated in certain models. While most eigenstates exhibit a volume-law entanglement entropy, a small subset obeys a sub–volume-law scaling.
These atypical eigenstates are known as quantum many-body scars.\jj{ They can give rise to long-lived memory of initial conditions and weak ergodicity breaking} \cite{Pizzi2025}.

As a paradigmatic example, we study the Kraus dynamics of the PXP model, whose Hamiltonian is given by
\begin{eqnarray}
H_{\rm PXP}=\frac{\Omega}{2}\sum_{m=0}^{n_H-1}P_{m-1}^0\sigma_m^xP_{m+1}^0 ~,
\end{eqnarray}
where $\Omega$ sets the overall energy scale and $P_m^0 = |0_m\rangle\langle 0_m|$ is a local projector. We impose open boundary conditions by taking $P^0_{-1}=P^0_{n_H}=1$.

The model features local constraints arising from the Rydberg blockade, encoded in the operators $P^0_m P^0_{m+1}$. These local conserved quantities commute with the superoperator $\mathcal{K}$. Without loss of generality, we therefore restrict our analysis to the constrained subspace satisfying $P^0_m P^0_{m+1}=1$. The resulting eigenvalue distribution is shown in Fig.~\ref{fig:PXP}(a). Compared to fully chaotic systems, which display a triangular spectral distribution, the spectrum here is noticeably smoother.

Importantly, we observe an enhanced population of eigenmodes at large $|\lambda|$ on the real axis compared to the purely chaotic case. The PXP model hosts four exact scar states that admit a matrix-product-state representation \cite{Lin2019}. To quantify their influence, we compute the average rescaled overlap between each eigenmode and  these scar states, $\xi_{\rm scar}=(\xi_1+\xi_2+\xi_3+\xi_4)/4$, shown in Fig.~\ref{fig:PXP}(b). While most eigenmodes cluster around $\xi_{\rm scar}\approx 1$, those with large $|\lambda|$ exhibit significantly enhanced overlap. This indicates that these eigenmodes are predominantly associated with scar states, which are consequently more robust against bath-induced resetting processes \cite{Zlatko2025}.

% \subsection{}
% \subsubsection{}

\begin{figure}[tb]
\includegraphics[clip = true, width =\columnwidth]{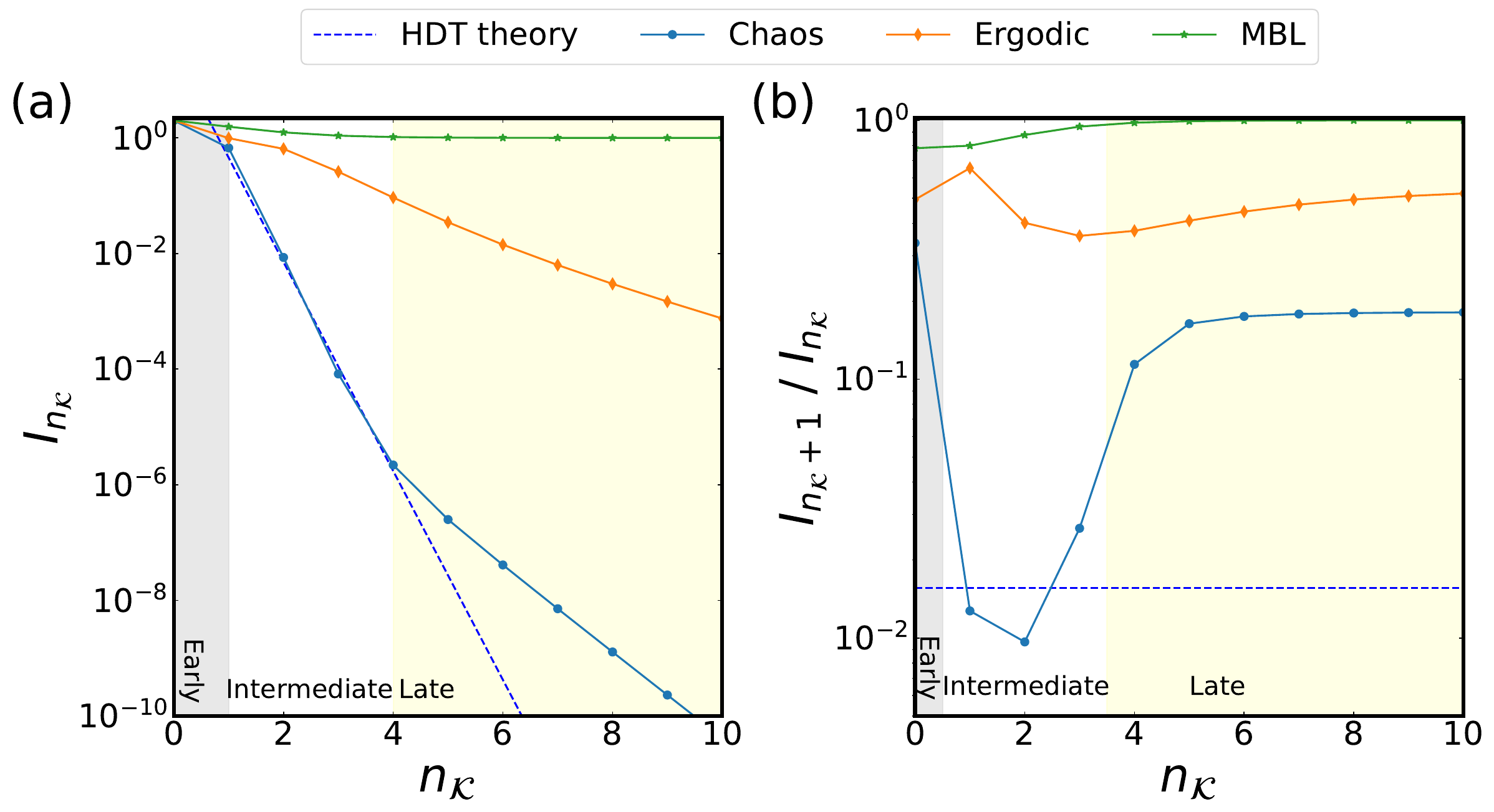}
\caption{\label{fig:QMI} Quantum mutual information as a function of the number of superoperations $n_\mathcal{K}$.
The system consists of $n_{\rm s}=6$ system qubits and $n_{\rm b}=6$ bath qubits.
The chaotic case is $J_{xxx}=2J_{xx}$ and others case are  $J_{xxx}=0$.
The MBL case is $J_z=5$ and other case are $J_z=0.1J_{xx}$. Other parameters are $J_{zz}=0.1J_{xx}$, $t=100\hbar/J_{xx}$ . The dashed line indicates the theoretical scaling $I\sim N_{\rm b}^{-n_\mathcal{K}}$.
}
\end{figure}

\section{Quantum Mutual Information and Correlation}

Quantum mutual information (QMI) provides a powerful diagnostic of information flow and mixing in quantum dynamics. In diffusive models, QMI is directly related to the convergence rate between ensembles, while in reservoir computing it characterizes the memory time of the dynamics. Experimentally, QMI can be accessed efficiently using shadow tomography.

A central motivation is to extract the eigenvalues of the dynamical map, which govern relaxation according to Eq. \ref{eq:Kn}. However, simple observables—such as local magnetization, two-point correlations, or state fidelity—generically exhibit Hamiltonian-dependent oscillations. These coherent effects obscure the asymptotic decay and hinder a reliable extraction of $\lambda_m$.

Going beyond local probes by considering entanglement within the system does not fully resolve this issue: entanglement can itself grow under unitary dynamics, and preparing highly entangled initial states across the entire system is experimentally challenging.

We therefore introduce a small reference system that is initially entangled with a subsystem and monitor the decay of the QMI. Owing to the data-processing inequality, QMI decays monotonically under the reduced dynamics, providing a smooth and robust probe of relaxation that enables a clean extraction of the spectral properties of the dynamical map.

We introduce a single ancilla qubit as a reference, which is jointly prepared with the system qubits in a GHZ state.
The QMI in the Rényi entropy form is
\begin{eqnarray}
    S=-\ln {\rm Tr}\left(\rho_{\rm a}^2\right)- \ln {\rm Tr}\left(\rho_{\rm s}^2\right)+\ln {\rm Tr}\left(\rho_{\rm as}^2\right)
\end{eqnarray}
The quantum mutual information after $n_\mathcal{K}$ applications of the operation $\mathcal{K}$ is shown in Fig.~\ref{fig:QMI}. For a chaotic Hamiltonian, the QMI exhibits the fastest decay. In an intermediate regime, the decay follows the scaling $S\sim N_{\rm b}^{-n_\mathcal{K}}$. At large $n_\mathcal{K}$ it becomes dominated by a small number of outliers with large eigenvalues. In the ergodic case, the decay is slower and never reaches the scaling $N_{\rm b}^{-n_\mathcal{K}}$. Although the QMI continues to decrease, the decay rate further slows down at large $n_\mathcal{K}$.
In contrast, for the MBL case, the QMI rapidly saturates and approaches a constant value in the large-$n_\mathcal{K}$ regime.

\begin{figure}[tb]
\includegraphics[clip = true, width =\columnwidth]{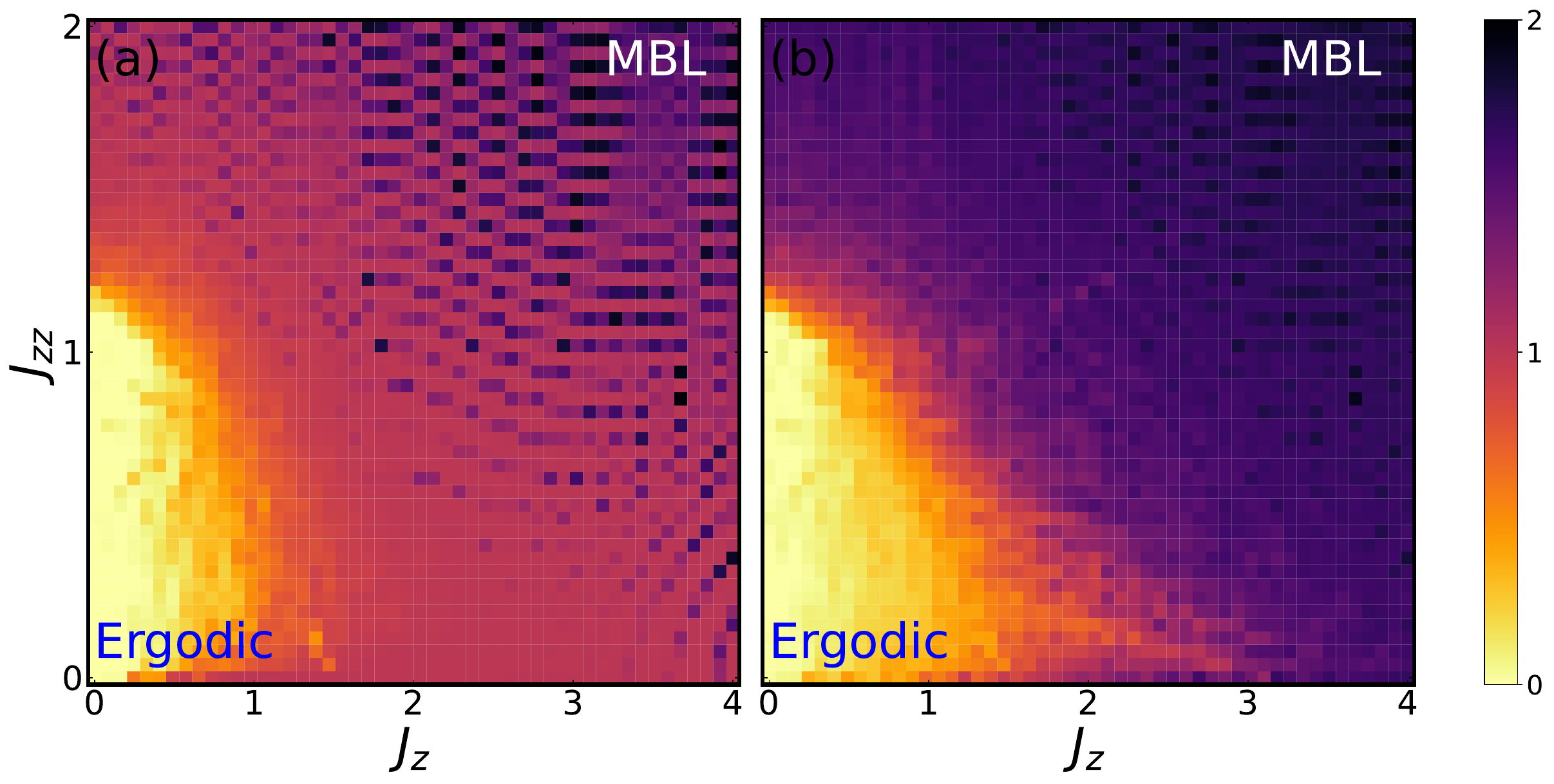}
\caption{\label{fig:Phase} (a) Mutual information $S$ and (b) imbalance $B+1$ as functions of the tuning parameter. Other parameters are $t=100\hbar/J_{xx}$, $n_{\mathcal{K}}=20$, $n_{\rm s}=6$, and $n_{\rm b}=10$.
}
\end{figure}

We further use the quantum mutual information (QMI) to distinguish dynamical phases. Figure~\ref{fig:Phase}(a) shows the parameter dependence of the QMI after 20 Floquet intervals: the ergodic regime exhibits a low QMI, whereas the MBL regime retains a high QMI.

For comparison, we also characterize the phases via the imbalance~\cite{Sierant2022,liu2026},
\begin{eqnarray}
B=\sum_i {\rm Tr}(\rho_t S_i^z){\rm Tr}(\rho_0 S_i^z)~,
\end{eqnarray}
as shown in Fig.~\ref{fig:Phase}(b). The imbalance yields a consistent phase boundary and agrees well with the QMI-based diagnosis in Fig.~\ref{fig:Phase}(a).
% \begin{figure}[tb]
% \includegraphics[clip = true, width =\columnwidth]{TC.pdf}
% \caption{\label{fig:TC} Jzz=0.1.
% }
% \end{figure}

%\section{Ergodic–MBL Transition}

\section{Results on IBM quantum devices}

\jj{To verify their dynamical behavior, we also executed the protocol on the IBM Quantum Platform. Quantum mutual information is challenging to measure in the current noisy intermediate-scale quantum (NISQ) era. Instead, we use the quantum-to-classical mutual information, whose expectation is defined as
\begin{equation}
	I_{\mathrm c} =\mathbb{E} D_{\mathrm{KL}}\left[P(z_{\mathrm a},z_{\mathrm s}) || P(z_{\mathrm a})\otimes P(z_{\mathrm s})\right]~,
\end{equation}
where $D_{\rm KL}$ is the Kullback-Leibler divergence. $P(\cdot)$ denotes the corresponding probability distribution. $z_{\rm a}$ and $z_{\rm s}$ are the measurement outcomes of the ancilla and system qubits, respectively.}

\jj{The initial GHZ state shared by the ancilla and system qubits is prepared using a cascade of CNOT gates. The Hamiltonian evolution is implemented through a Suzuki-Trotter decomposition. The bath reset is performed by a dynamic circuit: if the measurement outcome is 0, no operation is applied; if it is 1, an X gate is applied to the bath qubit to flip it to 0. Finally, the ancilla and system qubits are measured, and the probabilities of all measurement configurations are recorded.}

\begin{figure}[tb]
\includegraphics[clip = true, width =\columnwidth]{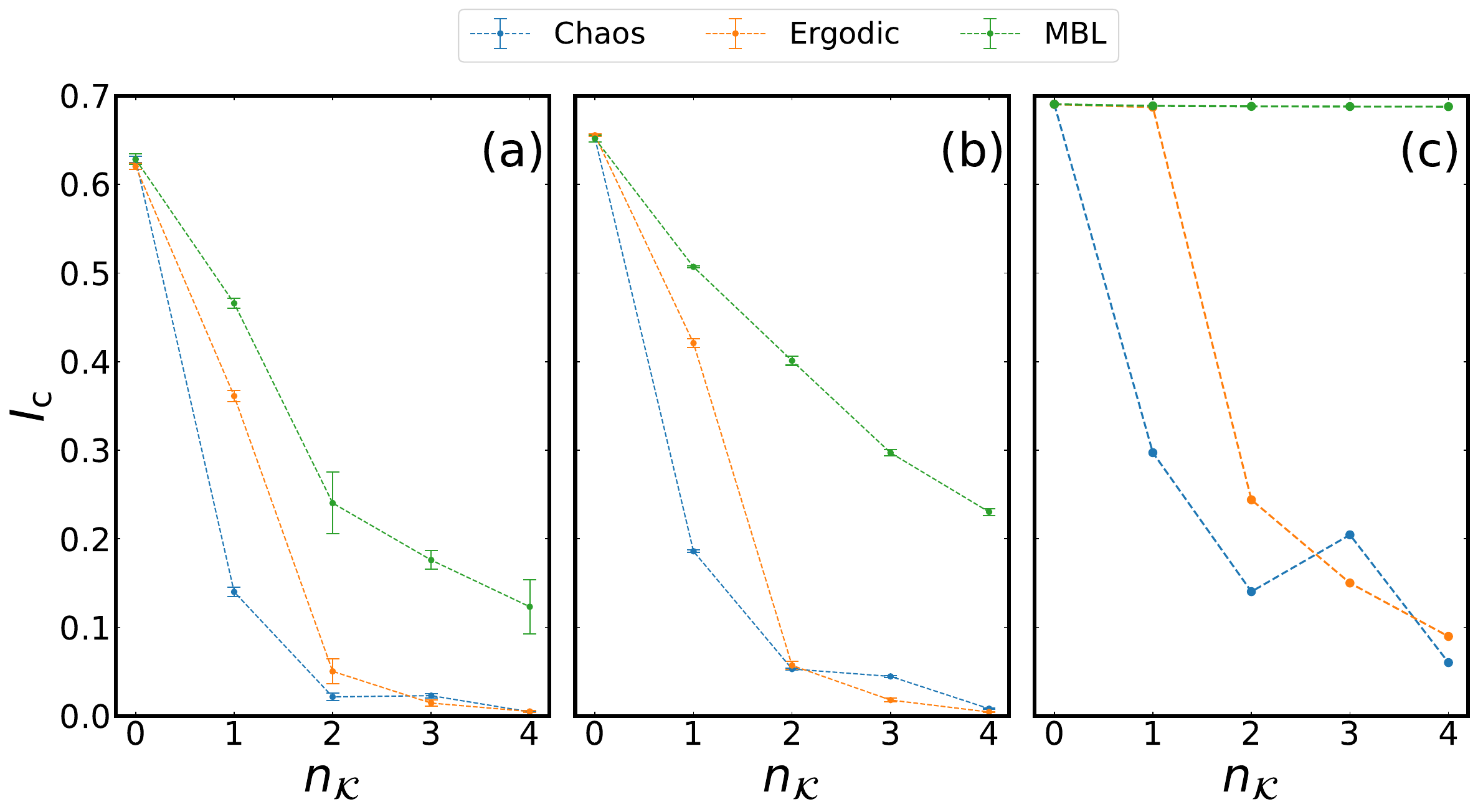}
\caption{\label{fig:Exp} Quantum-to-classical mutual information obtained from (a) the IBM Quantum platform, (b) noisy simulation, and (c) exact evolution. The numbers of ancilla, system, and bath qubits are $n_{\rm a}=1$, $n_{\rm s}=3$, and $n_{\rm b}=2$, respectively. The evolution time per iteration is $t=\hbar/J_{xx}$, and the number of Trotter steps is $r=3$. For the MBL case, $J_z=4J_{xx}$; for all other cases, $J_z=0.1J_{xx}$. For the chaotic case, $J_{yy}=0.5J_{xx}$; for all other cases, $J_{xx}=J_{yy}$.
}
\end{figure}

\jj{In Fig. \ref{fig:Exp}, we show the quantum-to-classical mutual information for different Hamiltonian parameters. In the MBL regime, the mutual information decays the slowest. The decay observed in the real device and noisy simulation [Figs. \ref{fig:Exp}(a) and \ref{fig:Exp}(b)] is mainly attributed to noise and errors, whereas in the noiseless case [Fig. \ref{fig:Exp}(c)] there is almost no decay. The decay is faster in the ergodic regime and fastest in the chaotic regime.}

\section{Conclusion and Outlook}
We have analyzed the eigenvalue problem of a quantum channel generated by Hamiltonian evolution and bath resetting. 
\jj{For the limited set of models studied here, the spectral structure of the channel differs qualitatively across chaotic, ergodic, and MBL regimes.} In the chaotic regime, the spectral density is well described by the circular law, accompanied by large-magnitude outlier eigenvalues that are correlated with low-energy properties of the underlying Hamiltonian. In the ergodic regime, the spectral density instead exhibits a smooth decay with increasing eigenvalue magnitude. In the MBL regime the spectrum becomes sparse and discrete. Notably, we identify eigenvalues with modulus close to unity. Such spectral features support long-lived period-multiplied dynamics and may be regarded as a spectral hallmark of discrete time-crystalline behavior. The decay of quantum mutual information further distinguishes these regimes, revealing different characteristic time scales for information relaxation and retention.

Beyond these spectral features, we have shown that the crossover between chaotic and ergodic channel dynamics is accompanied by the emergence of exceptional points. These non-Hermitian degeneracies signal the breaking of local dynamical symmetries of the channel and provide a complementary perspective on the transition between distinct open-system dynamical regimes.
%Taken together, our results demonstrate that the eigenvalue structure of reset-induced quantum channels encodes rich information about both the underlying Hamiltonian and the resulting open-system dynamics.
\jj{Taken together, for the models studied here, our results suggest that the eigenvalue structure of reset-induced quantum channels encodes information about both the underlying Hamiltonian and the resulting open-system dynamics. Whether this yields a general diagnostic across broader classes of many-body dynamics remains an open question.}

Several directions naturally follow from this work. First, while we have focused on representative classes of Hamiltonians, many structured or fine-tuned models remain to be explored. Establishing a systematic correspondence between Hamiltonian properties and the spectra of the associated reset-induced channels would provide a useful framework for classifying open many-body dynamics beyond conventional closed-system diagnostics. Second, our results suggest the possibility of engineering channel spectra through Hamiltonian control. Reset-driven Floquet channels, in particular, offer a natural platform for realizing robust subharmonic responses and discrete time-crystalline dynamics.

\begin{acknowledgements}
Q.Z. and J.J. acknowledge support from NSF (CCF-2240641, OMA-2326746, 2350153), ONR N00014-23-1-2296, AFOSR MURI FA9550-24-1-0349 and DARPA (HR0011-24-9-0362, HR00112490453, D24AC00153-02). Q.Z. also acknowledges support from ARPA-E (DE-AR0002067). The access to IBM Quantum Systems is provided
through USC’s IBM Quantum Innovation Center.
\end{acknowledgements}

\appendix

\section{Properties of Chaotic and Ergodic Systems \label{app:CE}}

\begin{figure}[tb]
	\includegraphics[clip = true, width =\columnwidth]{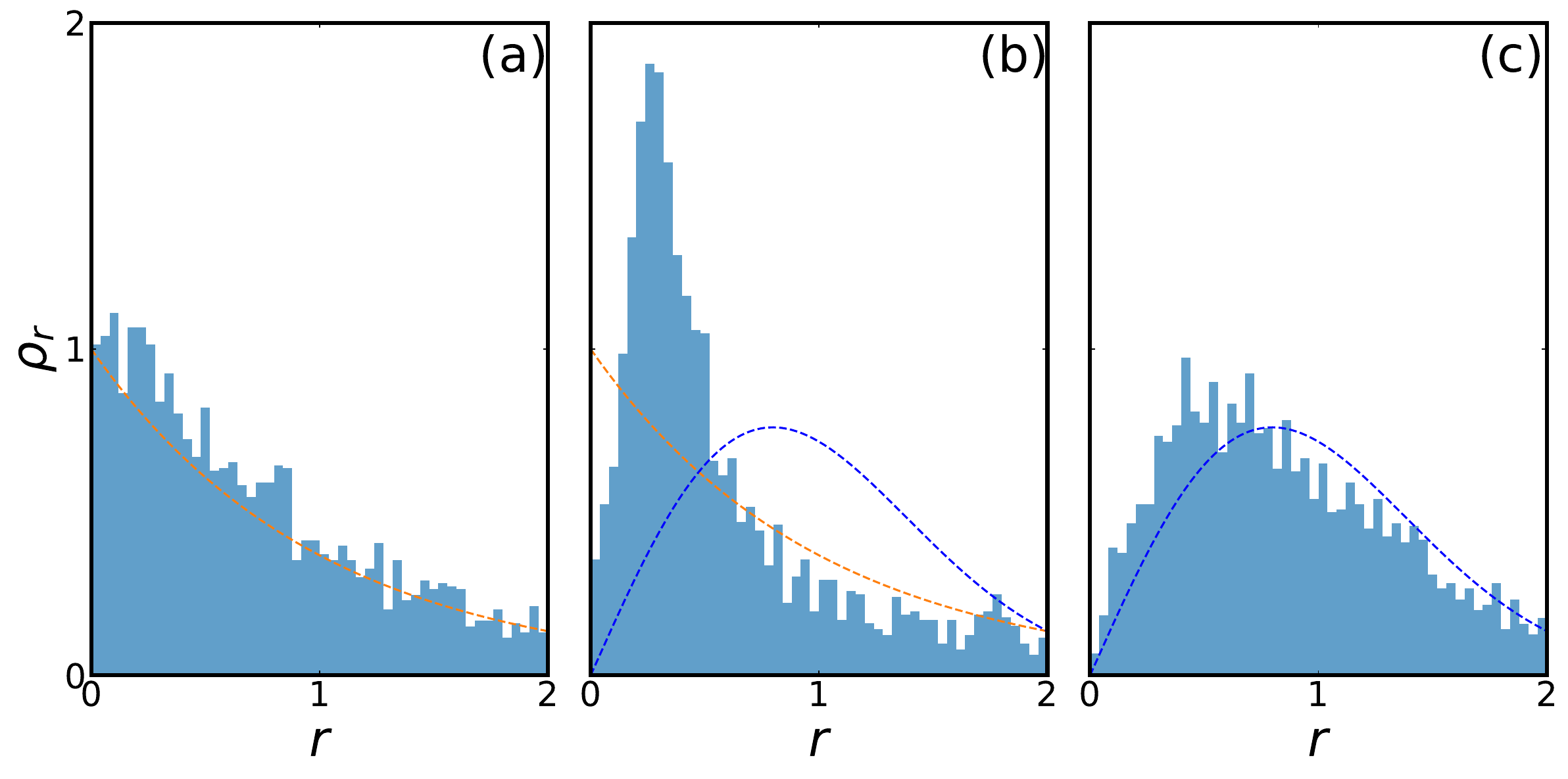}
	\caption{\label{fig:EigDis} Probability distribution of nearest-neighbor eigenenergy-level spacings for the Hamiltonian. The numbers of system and bath qubits are $n_{\rm s}=6$, and $n_{\rm b}=6$, respectively. Other parameters are $J_{z}=J_{zz}=0.1J_{xx}$. (b) $J_{yy}=0.3J_{xx}$; for all other cases, $J_{yy}=J_{xx}$. (c) $J_{xxx}=2J_{xx}$; for all other cases, $J_{xx}=J_{yy}$. The orange dashed line is the Poisson distribution and the blue dashed line is the Wigner-Dyson distribution. 
	}
\end{figure}

\jj{For a quantum chaotic system, the Hamiltonian matrix is well approximated by a random matrix \cite{Rodriguez2024,Kos2018}. Consequently, its dynamics are chaotic, its energy levels exhibit repulsion, and the nearest-neighbor energy-level spacing distribution obeys Wigner–Dyson statistics \cite{Rozenbaum2019}.}

\jj{However, since we are considering a physical system, its behavior is not identical to that of an ideal random matrix. Physical constraints, such as the locality of interactions, can give rise to a small number of quasi-conserved quantities \cite{Abdelshafy2025}. The number of such quantities is very small compared with the Hilbert-space dimension of the system.}

\jj{Although an ergodic system can be chaotic, for brevity we use “ergodic” in this manuscript to denote the nonchaotic case, i.e., systems that cannot be described by random-matrix theory. Such a system may possess many global conservation laws. Within each conserved subspace, the time average of an observable approaches its spatial average, and the nearest-neighbor energy-level spacing distribution obeys Poisson statistics \cite{Relano2004,Scaramazza2016}.}

\jj{To investigate nearest-neighbor level-spacing statistics, we sort the eigenenergies in ascending order and define the dimensionless nearest-neighbor spacing
\begin{eqnarray}
	r_n=\frac{E_{n+1}-E_n}{ \frac{E_{N}-E_1}{N-1}  } ~,
\end{eqnarray}
where $E_n$ is the $n$th eigenenergy and $N$ is the dimension of the Hilbert space.}

\jj{In Fig. \ref{fig:EigDis}, we show the nearest-neighbor energy-level spacing distributions for different Hamiltonian parameter sets. To improve the quality of the fits, we exclude the lowest and highest 10\% of eigenstates, whose physical properties, such as entanglement entropy, differ from those of the bulk states. In Fig. \ref{fig:EigDis}(a), we show the ergodic case. The spacing distribution is well fitted by a Poissonian distribution. By tuning $J_{yy}$, we break the $U(1)$ symmetry, which introduces weak chaos. As shown in Fig. \ref{fig:EigDis}(b), the resulting spacing distribution lies between the Poissonian and Wigner–Dyson distributions. Although the system is not fully chaotic, we use this case as a representative chaotic example in the experiment because it is easier to realize on a quantum computer. Finally, adding a three-body interaction term $J_{xxx}$, the system can exhibit strong chaos as shown in  Fig. \ref{fig:EigDis}(c). It can be well fitted by the  Wigner–Dyson distribution.}

\section{Eigenstructure at EPs \label{app:EP}}

At EPs, the channel becomes defective, and the eigenoperator expansion used in the main text is no longer complete. The resulting singularity requires a generalized treatment in operator space. In this section, we discuss how to handle this case by introducing the appropriate Jordan-chain structure of the channel.

The channel written in Kraus form is
\begin{eqnarray}
    \mathcal{K}\left(X\right)=\sum_\alpha K_\alpha X K_\alpha^\dagger
\end{eqnarray}
with eigenoperator $\varrho_m$ and eigenvalue $\lambda_m$
We can define its dual channel
\begin{eqnarray}
    \mathcal{K}^\dagger\left(X\right)=\sum_\alpha K_\alpha^\dagger X K_\alpha
\end{eqnarray}
with eigenoperator $\mu_m$ and eigenvalue $\lambda_m^*$

At the EPs, the basis is incomplete. We can use the generalized eigenoperators to compensate for it
\begin{eqnarray}
    \mathcal{K}\left(\varrho_m^{(0)}\right)-\lambda_m\varrho_m^{(0)}&=&0 ~,\nonumber\\
    \mathcal{K}\left(\varrho_m^{(d)}\right)-\lambda_m\varrho_m^{(d)}&=&\varrho_m^{(d-1)} ~(1\leq d \leq o-1)
\end{eqnarray}
where $o$ is the order of the corresponding EP.

Decomposing the initial state in the eigenoperator basis as
\begin{eqnarray}
    \rho_{\rm s0}=\sum_{m}c_m^{(d)} \varrho_m^{(d)} ~,
\end{eqnarray}
where the coefficient is
\begin{eqnarray}
    c_m^{(d)}={\rm Tr} ({\mu_m^{(d)}}^\dagger\varrho_m^{(d)}) ~.
\end{eqnarray}
For $n_{\rm r}$ repeated applications of the quantum channel, the output is
\begin{eqnarray}
    \mathcal{K}^{n_{\rm r}}\left(\rho_{\rm s0}\right)=\sum_m \sum_{d=0}^{o-1} \sum_{\Delta d=0}^{\min (o-1-d,n_{\rm r})} \frac{n_{\rm r}!}{(n_{\rm r}-\Delta d)!\Delta d!} \lambda_m^{n_{\rm r}-\Delta d} c_m^{(d+\Delta d)} \varrho_m^{(d)} ~. \nonumber\\
\end{eqnarray}

%\section{Anisotropic Heisenberg Model}
%
%We next consider a more general anisotropic case with unequal coupling coefficients. This anisotropy breaks the conservation of total spin and drives the Hamiltonian into a chaotic regime. The model is given by
%\begin{eqnarray}
%H_{\rm XX}&=&\sum_{m=0}^{n_{H}-2} \left( J_{xx} \sigma_m^x \sigma_{m+1}^x+J_{yy} \sigma_m^y\sigma_{m+1}^y+J_{zz}\sigma_m^z\sigma_{m+1}^z\right)\nonumber\\
%&&+J_z\sum_{m=0}^{n_{H}-1}\cos(\omega m) \sigma_m^z ~.
%\label{eq:XX}
%\end{eqnarray}
%
%As shown in Fig.~\ref{fig:EP_XX}, when $J_{xx}$ deviates from $J_{yy}$, the system encounters more EPs, and the number of complex eigenvalues increases correspondingly.
%
%\begin{figure}[tb]
%\includegraphics[clip = true, width =\columnwidth]{EP_XX.png}
%\caption{\label{fig:EP_XX}
%Selected eigenvalue bands of $\mathcal{K}$ with large positive real parts. (a) Imaginary parts of the eigenvalues as a function of the chaos parameter. Prisms mark representative EPs. (b) Number of eigenvalues with a nonzero imaginary part. Other parameters are $J_{zz}=J_z=0.1$, $t=1000\hbar/J_{yy}$, and $n_{\rm s}=n_{\rm b}=5$.
%}
%\end{figure}

%\begin{acknowledgments}
%\end{acknowledgments}

\bibliography{Decay}

\end{document}